\documentclass[
    fontsize=11pt,DIV=10,fleqn,
    letterpaper,oneside,paper=portrait,
    headings=standardclasses
]{scrartcl}

\usepackage[scaled=.98,sups,osf]{XCharter}
\usepackage[scaled=1.04,varqu,varl]{inconsolata}
\usepackage[type1]{sourcesanspro}
\usepackage[uprightscript,libertine,vvarbb,scaled=1.05]{newtxmath}

\usepackage{scrlayer-scrpage}
\usepackage{mathtools}
\usepackage{bm}
\usepackage[eqno,enum,lineno]{tabfigures}
\usepackage{setspace}
\usepackage{ragged2e}
\usepackage{booktabs}
\usepackage{ntheorem}
\usepackage[backend=biber,authordate,noibid]{biblatex-chicago}
\usepackage{footmisc} 
\usepackage[allbordercolors={0.8 0.8 1}]{hyperref}
\usepackage[nameinlink,noabbrev]{cleveref}

\usepackage[para,flushleft]{threeparttable}
\usepackage{dcolumn}
\usepackage[inline]{enumitem}

\usepackage{tabulary}
\usepackage{subfig}
\usepackage{tikz}
\usepackage{pgfplots}
\usepackage[outline]{contour}
\usepackage{multicol}

\KOMAoptions{onpsinit={\linespread{1}\selectfont}}

\allowdisplaybreaks[1]

\newtheorem{hypothesis}{Hypothesis}

\DeclareCiteCommand{\citeyearlinked}
  {\boolfalse{citetracker}%
   \boolfalse{pagetracker}%
   \usebibmacro{prenote}}
  {\ifciteindex
     {\indexfield{indextitle}}
     {}%
   \printtext[bibhyperref]{\printfield[citeyear]{labelyear}}}
  {\multicitedelim}
  {\usebibmacro{postnote}}

\newcounter{savefootnote}
\newcounter{symfootnote}
\newcommand{\symfootnote}[1]{%
  \setcounter{savefootnote}{\value{footnote}}%
  \setcounter{footnote}{\value{symfootnote}}%
  \ifnum\value{footnote}>8\setcounter{footnote}{0}\fi%
  \let\oldthefootnote=\thefootnote%
  \renewcommand{\thefootnote}{\fnsymbol{footnote}}%
  \footnote{#1}%
  \let\thefootnote=\oldthefootnote%
  \setcounter{symfootnote}{\value{footnote}}%
  \setcounter{footnote}{\value{savefootnote}}%
}

\newcolumntype{d}[1]{D{.}{.}{#1}}

\pgfplotsset{compat=1.16}

\usetikzlibrary{calc}

\contourlength{2pt}

\makeatletter
\def\input@path{{../tables/}}
\makeatother
\graphicspath{{../figures/}}
\begin{document}

\hphantom{.}
\vfill

\begin{center}

\textbf{\LARGE Temptation: Immediacy and certainty\symfootnote{I give special thanks to Gary Charness, Ignacio Esponda, and Erik Eyster. I also thank David Freeman, Peter Kuhn, and participants at University of California, Santa Barbara, seminars and Economic Science Association meetings. \textcite{Reddinger2019} pre-registered this study with the American Economic Association as AEARCTR-0004651. The University of California, Santa Barbara, Human Subjects Committee exempted my protocol 56-19-0621. \citeauthor{Reddinger2024} (\citeyearlinked{Reddinger2024}) provides data and source code. The Department of Economics and the Graduate Division at the University of California, Santa Barbara, provided funding.}}

\vfill

{\large

J.~Lucas Reddinger\symfootnote{Department of Economics, University of California, Santa Barbara. Current affiliation: Department of Economics, Purdue University, 403 West State Street, Krannert Building, Room 712, West Lafayette, Indiana 47907-2076; \href{mailto:reddinger@purdue.edu}{reddinger@purdue.edu}.}


\vspace{\baselineskip}

26 August 2024

}

\end{center}

\vfill
\vspace{-1\baselineskip}

\paragraph{Abstract}
Is an option especially tempting when it is both immediate and certain?  I test the effect of risk on the present-bias factor given quasi-hyperbolic discounting.  In my experiment workers allocate about thirty to fifty minutes of real-effort tasks between two weeks.  I study dynamic consistency by comparing choices made two days in advance of the workday with choices made when work is imminent.  My novel design permits estimation of present bias using a decision with a consequence that is both immediate and certain.  I find greater present bias when the consequence is certain.  This finding has implications for any economic decision involving a present-biased decision-maker, including labor contracting and consumer good pricing.  I offer a methodological remedy for experimental economists.

\vfill
\vspace{-1\baselineskip}

{\RaggedRight

\paragraph{\emph{JEL} Codes} C91, D80, D90

\paragraph{Keywords} present bias, dynamic inconsistency, quasi-hyperbolic discounting, time preferences, risk preferences, immediacy effect, certainty effect, experimental economics

}

\vfill

\clearpage

\spacing{1.5}

\section{Introduction}

While risk and time preferences are fundamental to the theory of decision-making, much remains unknown about the interplay between these two dimensions.  A future prospect is inherently risky if any circumstance may arise that precludes consumption of the consequence.  This implies that outcomes are obtained with certainty only if obtained without delay.  Accordingly a preference for certain outcomes results in a preference for immediate outcomes.  Conversely, the introduction of risk may especially diminish the appeal of an immediate reward, relative to a delayed reward.  I explore in this paper such an interaction between immediacy and certainty with an experiment of dynamic decision-making over risky and delayed prospects.

Present-biased preferences explain oft-bemoaned consumer behavior, such as the failure to meet one's own physical exercise goals and the over-utilization of credit-card debt \parencite{RoyerStehrSydnor2015,MeierSprenger2010}.  Firms exploit consumer present bias and successfully extract welfare \parencite{DellaVignaMalmendier2004}.  Meanwhile, incentives, commitment devices, and other interventions may (or may not) help consumers improve their long-term welfare \parencite{AshrafKarlanYin2006,CarreraRoyerStehrSydnorTaubinsky2022}.  A better understanding of present bias assists this body of research.

As an example, my study informs labor contract design, especially those used in the modern gig economy.  Consider drivers for ride-hail companies---these workers face decisions similar to those of the workers in my experiment.  Ride-hail companies carefully withhold selective information (such as ride length or destination) when offering a gig to a driver and require commitment prior to revealing all of these ride details \parencite{Rana2020}.  Such uncertainty in a spot labor contract theoretically affects labor supply; my results confirm that a present-biased worker with a weekly income target may procrastinate less given greater uncertainty.

My novel experimental design allows estimation of present bias for subjects making a single decision with a certain consequence.  This is in contrast to a baseline treatment that implements an allocation choice made on a randomly-selected day and at a randomly-selected intertemporal price ratio, in accord with prevailing experimental methodology.

Workers in my experiment allocate a workload between two weeks.  Each worker first makes allocation decisions two days before the first workday.  Each worker then returns on the first workday and makes identical decisions with the work being imminent.  If a worker is present-biased, she will in advance choose some allocation between the two weeks, but then on the first workday, prefer an allocation with less work for the present day.  In my experiment, once the implementation mechanism selects a particular allocation, the worker must complete the tasks allocated to each week to earn a substantial bonus payment.

Ultimately I find that the immediacy effect is significantly attenuated by the introduction of risk.  Specifically, the quasi-hyperbolic present-bias factor $\beta$ becomes smaller with the elimination of risk, implying greater myopia.  In my experiment, when a workload allocation is implemented with certainty, subjects on average discount the future by a factor of $\hat\beta=0.581$ relative to the present.  In the baseline treatment that uses prevailing experimental methodology, I find no statistically significant present bias, with $\hat\beta=1.009$ when each decision has a 10\% implementation probability.

These findings underscore the importance of decision-theoretic frameworks that permit interaction between dimensions of risk and time.  The findings also suggest that studies of tempting goods may necessitate decisions with temporally salient and certain consequences; researchers should keep this in mind when designing either lab or field experiments.  Further, risk introduced by randomized incentive mechanisms---common experimental methodology---may require augmentation of decisions with certain consequences.  I offer such a methodological remedy.

\section{Background} \label{sec:background}

To model intertemporal decision-making, \textcite{Samuelson1937} introduced \emph{exponential discounted utility} (DU), which describes how an individual values utility flows (of consumption goods, such as leisure) that occur over time. If utility flows $u(x_{t+\tau})$ result from consumption $x_{t+\tau}$ at time $t+\tau \in \mathbb{N}$, given a constant discount factor $\delta \in [0,1]$, the model gives an intertemporal value at time $t$ of%
\begin{align}%
U^\textrm{DU}_t &= \sum_{\tau=0} \delta^\tau u(x_{t+\tau}).\label{eq:du}%
\end{align}%
A decision-maker with this value function will make \emph{dynamically consistent} choices, assuming that the felicity function $u$ is time-invariant \parencite{Halevy2015}.

To capture a preference for immediate utility, \textcite{Laibson1997} introduces the present-bias factor $\beta \geq 0$ to discount all future utility flows contra present utility.  The resultant \emph{quasi-hyperbolic discounted utility} (QHD) has an intertemporal value at time $t$ of%
\begin{align}%
U^\textrm{QHD}_t &= u(x_t) + \beta \sum_{\tau=1} \delta^\tau u(x_{t+\tau}).\label{eq:qhd}%
\end{align}%
$\beta<1$ describes a preference for immediacy, also referred to as \emph{present bias}. This is an example of \emph{diminishing sensitivity to delay}---an individual is more impatient regarding a delay in felicity that happens immediately relative to a delay that occurs in the future.  Meanwhile, some individuals may exhibit future bias, with $\beta>1$.

A decision-maker with $\beta\ne1$ will make \emph{dynamically inconsistent} choices. A decision-maker with $\beta<1$ ($\beta>1$) will continually revise consumption plans to achieve greater (lesser) felicity in the present moment relative to her prior plans.

\subsection{Diminishing sensitivity to risk and delay}\label{sec:background-risk-delay}

Let a simple gamble $(x \circ p)$ be a prospect that obtains $x$ with probability $p$. Given the independence axiom of expected utility theory (EUT), a preference relation is maintained if prospect probabilities are multiplied by a common ratio \parencite{Machina1982}. Consider%
\begin{align*}%
\text{Menu $\mathcal{A}$:} && a &= ( 1 \circ 0.9 ) && \text{or} & a' &= ( 2 \circ 0.6 ); & \text{and} \\*
\text{Menu $\mathcal{B}$:} && b &= ( 1 \circ 0.6 ) && \text{or} & b' &= ( 2 \circ 0.4 ). &
\end{align*}%
Under EUT, $a$ is weakly preferred to $a'$ if and only if $b$ is weakly preferred to $b'$.

\textcite{Allais1953} presented empirical violations of this result \parencite[see also][]{KahnemanTversky1979}.  The \emph{common ratio effect} describes a preference reversal in which a decision-maker is indifferent between two prospects, but when the prospect probabilities are scaled down by a common ratio, she then strictly prefers the riskier option (\emph{e.g.}, $a \sim a'$ and $b \prec b'$).

The \emph{certainty effect} is a special case of the common ratio effect when one prospect obtains with probability one. For example, consider%
\begin{align*}%
\text{Menu $\mathcal{C}$:} && c &= ( 3 \circ 1.0 ) && \text{or} & c' &= ( 4 \circ 0.8 ); & \text{and} \\*
\text{Menu $\mathcal{R}$:} && r &= ( 3 \circ 0.5 ) && \text{or} & r' &= ( 4 \circ 0.4 ). &
\end{align*}%
If a decision-maker is indifferent between $c$ and $c'$ but strictly prefers $r'$ over $r$, she may simply possess diminishing sensitivity to risk as described by the common ratio effect, or she may have a disproportionate preference for a certain outcome.

\textcite{PrelecLoewenstein1991} note equivalent results regarding time delay using discounted utility.  Following \textcite{Halevy2008}, let us simply interpret $\delta$ in \cref{eq:du} as a failure risk imposed by a unit-time delay that precludes consumption of the consequence (\emph{e.g.}, $\delta$ might be one's probability of death in every time period).  Under DU, a decision-maker weakly prefers one intertemporal consumption plan to another if and only if this preference is maintained with an additional arbitrary time delay.

Consider a daily survival probability of $0.8$. Let us reinterpret the previous menus as%
\begin{align*}%
\text{Menu $\mathcal{\tilde{C}}$:} && C &= 3 \text{ now} && \text{or} & C' &= 4 \text{ in } 1 \text{ day;} & \text{and}  &\\*
\text{Menu $\mathcal{\tilde{R}}$:} && R &= 3 \text{ in } 3 \text{ days} && \text{or} & R' &= 4 \text{ in } 4 \text{ days.\footnotemark}
\end{align*}%
\footnotetext{With approximation, 0.5 is the probability of surviving $\ln 0.5 / \ln 0.8 \approx 3$ days, and 0.4 is the probability of surviving $\ln 0.4 / \ln 0.8 \approx 4$ days.}%
The \emph{common difference effect} describes a preference reversal in which a decision-maker is indifferent between two intertemporal consumption plans, but when an arbitrary time delay is added to each, she then becomes more patient (\emph{e.g.}, $C \sim C'$ and $R \prec R'$).

The \emph{immediacy effect} is a special case of the common difference effect, when only immediate consumption varies between plans.  In \cref{eq:qhd}, QHD describes an immediacy effect if $\beta<1$. \textcite{ChakrabortyHalevySaito2020} fully characterize the relationship between the immediacy effect (including under QHD) and the certainty effect.\footnote{ \textcite{EpperFehrDuda2018,BaucellsHeukamp2010,GreenMyerson2004} also explore this relationship.\label{note:risk-delay-papers}}

In this sense, a decision-maker only obtains a prospect with certainty when also obtained without delay.  Any time delay plausibly eliminates a certainty effect, and similarly any risk plausibly eliminates an immediacy effect. I endeavor to experimentally test the significance of this interaction.

\subsection{Evidence of risk moderating present bias}

At least three studies have shown that risk moderates present bias using hypothetical or nearly-hypothetical monetary incentives.

\textcite{KerenRoelofsma1995} conduct a between-subject full-factorial experiment with hypothetical monetary stakes, wherein subjects choose between \$50 and \$55 with a four-week delay.  When prizes obtain with certainty, 82\% of subjects prefer \$50 immediately over \$55 in four weeks, while only 37\% of subjects prefer \$50 in twenty-six weeks over \$55 in thirty weeks, thereby demonstrating present bias at certainty. When prizes obtain with probability one-half, 39\% of subjects prefer \$50 immediately over \$55 in four weeks, and 33\% of subjects prefer \$50 in twenty-six weeks over \$55 in thirty weeks, failing to show significant present bias.

\textcite{WeberChapman2005} confirm these findings, again with hypothetical monetary stakes. \textcite{BaucellsHeukamp2010} also find that risk moderates present bias with highly-diluted monetary incentives, implementing only three of 3,757 decisions.\footnote{Three of 221 subjects are selected, each of whom has one of their seventeen decisions implemented.}

However the methodologies employed may not appropriately identify the effect of interest.  Hypothetical decisions lack incentive, relying solely on framing and contingent reasoning.  Similarly, extremely low probabilities of implementation may dilute the stated probability of the prospects due to isolation failure (which I discuss in \cref{sec:background-decision}).  Finally, monetary earnings do not necessarily translate to consumption as in the intertemporal models of \cref{eq:du,eq:qhd}; an individual would need to be extremely liquidity-constrained to trade a few dollars of earnings for a good to be consumed the same day.

Models of present bias are often used to study self-control failure of visceral urges \parencite{CheungTymulaWang2021}, which are plausibly best elicited with an immediate and certain consequence.  For example, many studies of sequential games find costly punishment more prevalent upon eliciting a direct-response action instead of a conditional strategy \parencite{BrandtsCharness2011}, perhaps due to a preference for exacting \emph{unconditional} (\emph{i.e.}, certain) revenge.

My study is the first to use truly immediate and certain consequences in studying present bias.  I avoid concerns associated with hypothetical decisions, long-shot implementation,  and monetary stakes, thereby establishing an ideal method for capturing present bias.

\subsection{Empirical estimates of present bias} \label{sec:background-estimates}

While many studies have used monetary rewards to measure present bias, \textcite{AndreoniSprenger2012a} pioneered the ``convex time budget'' (CTB) methodology, which elicits monetary-prize allocations between two time periods at various interest rates, thereby allowing risk preferences and QHD parameters to be estimated jointly.\footnote{See \textcite{CheungTymulaWang2021} for a nice review of estimates of present bias using various methodologies.  For more on these other methodologies used to elicit time and risk preferences, see \textcite{AndersenHarrisonLauRutstroem2008,Cheung2016,Cheung2020,AbdellaouiBleichrodtParaschiv2013}.}  However, monetary earnings may not adequately capture consumption utility in the absence of liquidity constraints and decision isolation.  \textcite{AugenblickNiederleSprenger2015} address this concern with CTB decisions in which individuals allocate real-effort tasks across time (other studies have used alternative primary rewards, such as food).

The meta-analysis by \textcite{ImaiRutterCamerer2021} finds no evidence of present bias in monetary rewards, while finding a mean bias-corrected present-bias factor $\beta$ between 0.90 and 0.99 in real-effort tasks.  While the specific value depends on the particular bias correction, present bias is also highly context-dependent.\footnote{While \textcite{ImaiRutterCamerer2021} restrict their focus to twenty-eight studies that use the CTB methodology, \textcite{CheungTymulaWang2021} offer a meta-analysis which includes studies that use other methodologies such as the joint-elicitation methodology \parencite{AndersenHarrisonLauRutstroem2008}.  Xueting Wang has informed me that a forthcoming revision of \textcite{CheungTymulaWang2021} will present results that are qualitatively consistent with those of \textcite{ImaiRutterCamerer2021}.}

\subsection{Decision framing, isolation, and implementation}\label{sec:background-decision}

A typical subject in an economics experiment makes many decisions.  Historically many experiments implement many or all decisions, but this method can yield data rife with wealth effects, hedging, and other confounds \parencite{CharnessGneezyHalladay2016}.  Consequently most experiments now implement one randomly-selected decision, thus avoiding such complementarity between outcomes \parencite{AzrieliChambersHealy2018}. Yet implementing one decision at random is not a panacea; many subjects still fail to isolate each decision.\footnote{See  \textcite{StarmerSugden1991,BeattieLoomes1997,CubittStarmerSugden1998,CoxSadirajSchmidt2015}.}

Non-expected utility rationalizes isolation failure when a subject views a set of decisions as comprising a single optimization problem.  For example, given a \textcite{HoltLaury2002} choice list, a subject is often able to secure a certain outcome by choosing every safe option.  \textcite{FreemanHalevyKneeland2019} find evidence of the certainty effect when comparing pairwise choices to choice-list data. \textcite{FreemanMayraz2019} find evidence of the Allais paradox with the effect independent of the mechanism used.

\textcite{FreemanMayraz2019} find that presentation has the largest impact on isolation. \textcite{BrownHealy2018} display decisions separately and reclaim incentive compatibility.

\section{Experimental design}

The present study compares present bias between a risky consequence and a certain consequence.  I implement one decision to avoid complementarity (see \cref{sec:background-decision}), which must be implemented with certainty in the respective treatment.  To obtain useful choice data from a single implemented decision, I use the CTB methodology (see \cref{sec:background-estimates}).  To induce an immediate (primary) reward, I ask subjects to allocate a budget of real-effort tasks between two weeks.

The experiment consists of three sessions: Monday (day zero), Wednesday (day two), and the following Wednesday (day nine).  Every subject earns \$1.50 per session, which must be completed between noon and midnight.  I immediately disqualify subjects who miss a deadline.  Every subject that completes all three sessions earns a \$5 bonus.

\begin{figure}[p]
\begin{multicols}{2}
\noindent\subfloat[Task interface\label{fig:ui-task}]{\fcolorbox{gray}{white}{\includegraphics[width=\columnwidth,trim=0 12cm 0 0, clip]{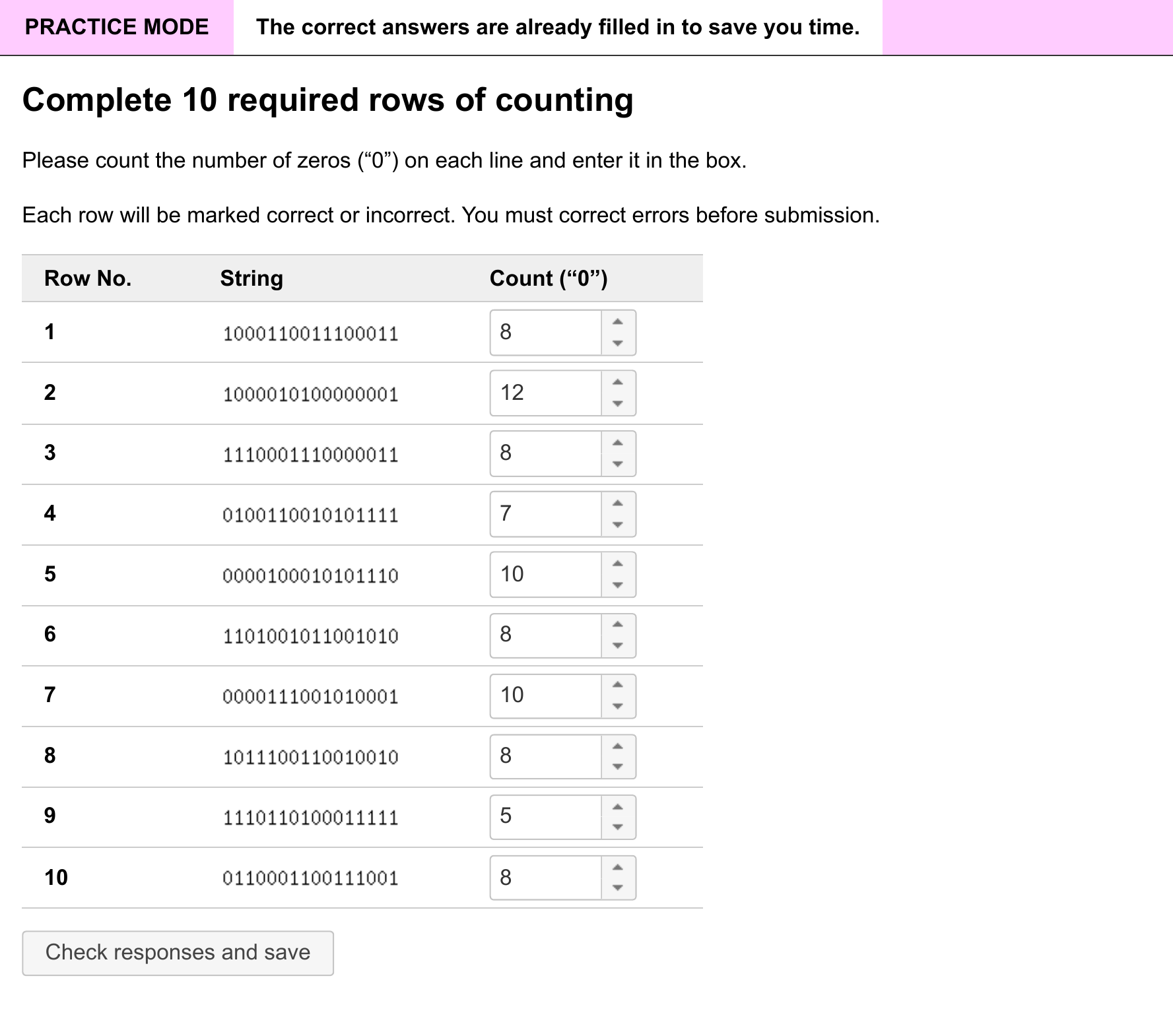}}}

\noindent\subfloat[Allocation interface, presented separately\label{fig:ui-allocation-a}]{\fcolorbox{gray}{white}{\includegraphics[width=\columnwidth,page=9]{figure_interface_3_demo_allocate.pdf}}}

\columnbreak

\noindent\subfloat[Allocation interface, presented juxtaposed\label{fig:ui-allocation-b}]{\fcolorbox{gray}{white}{\includegraphics[width=\columnwidth,trim=0 12cm 0 0, clip,page=17]{figure_interface_3_demo_allocate.pdf}}}

\noindent\subfloat[Day selection mechanism interface\label{fig:ui-mechanism}]{\centering\fcolorbox{gray}{white}{\includegraphics[width=\columnwidth]{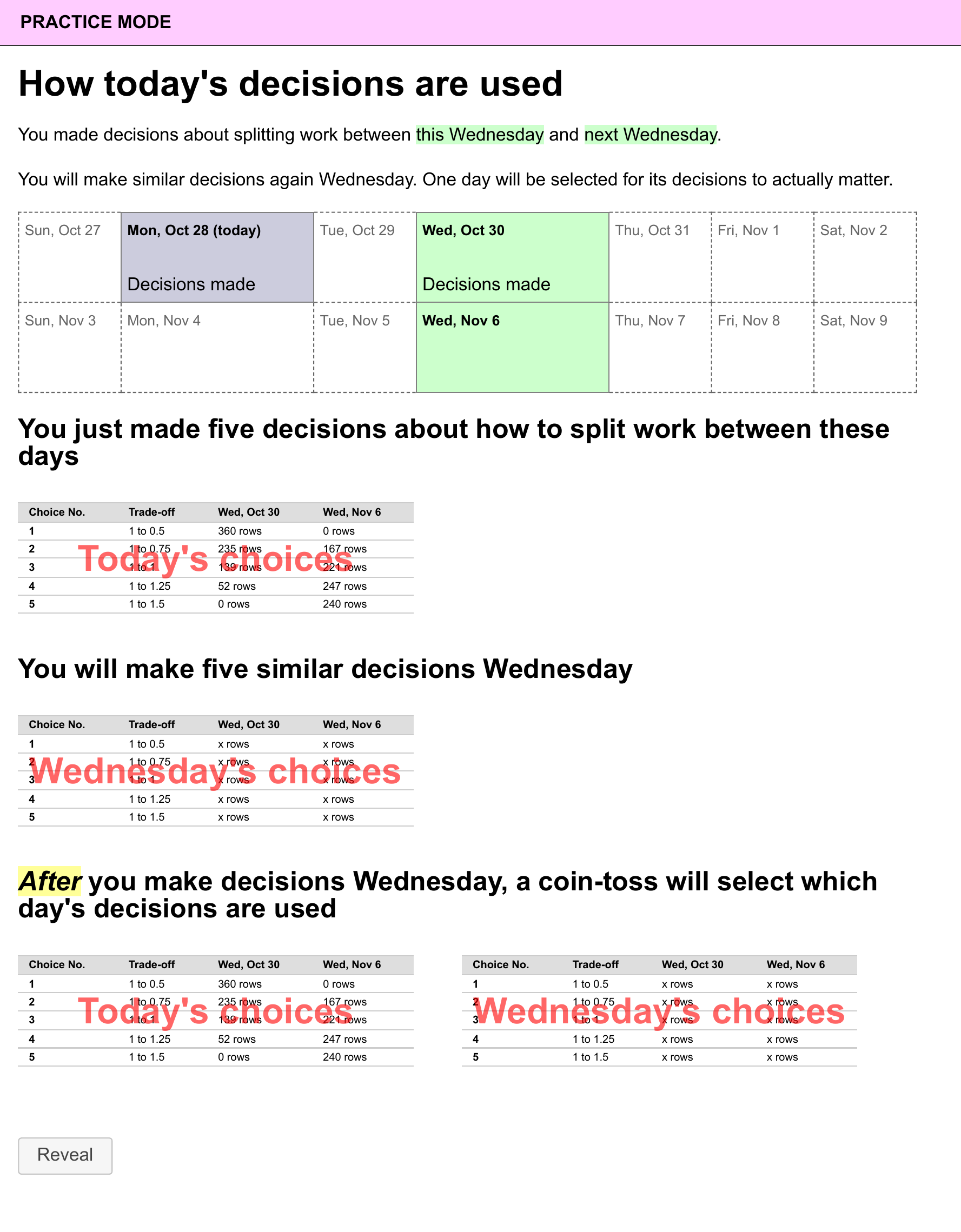}}}
\end{multicols}
\caption{Experimental interface\label{fig:interface}}
\end{figure}

Each session begins with ten mandatory tasks, providing salient experience and a fixed baseline effort-level on each day.  Each task asks the subject to count the number of zero digits in a sixteen-digit binary string and enter this count into an adjacent text field (\cref{fig:ui-task}).  The subject must remedy any incorrect response for successful submission.

Following the mandatory tasks on day zero and day two, each subject chooses an allocation of 360 tasks between day two and day nine at each of five substitution rates (the 360-task budget has day-two value).  Each decision is presented individually to elicit a tentative choice (\cref{fig:ui-allocation-a}), which are then juxtaposed on a single page (\cref{fig:ui-allocation-b}) for any adjustment.

Accordingly, let $e^t_{i,d}$ denote effort chosen at rate $R_i$ on decision-day $d$ to be expended on workday $t$. For example, $e^\textrm{day two}_{4,0}$ is the effort chosen at rate $R_4$ on day zero to be worked on day two.  Each subject faces the constraint $$e^\textrm{day two}_{i,d} + R_i e^\textrm{day nine}_{i,d} = 360, \text{ for each } R_i \in \mathcal{R} \coloneqq \left \langle 1.25, 0.75, 1, 1.5, 0.5 \right \rangle \text{ and } d \in \left \{0, 2 \right \} .$$ Some rates $R_i$ (which can also be interpreted as productivity ratios or gross interest rates) entail substantial income effects. For example, at $R_4=1.5$, a subject may choose to delay all 360 tasks on day two to only 240 tasks on day nine. Yet if a subject delays all 360 tasks at $R_5=0.5$, she would need to complete 720 tasks on day nine.  Randomly-selected subjects receive a rate sequence in reverse order.

\subsection{Treatments}

I implement a $2 \times 2$ factorial between-subject design. I inform all subjects that a decision from either day zero or day two will be selected with equal probability (\cref{fig:ui-mechanism}). \nocite{BrandtsCharness2011}

The baseline treatment implements one of the ten decisions with uniform probability.  Subjects with this treatment have uncertainty regarding the day and the rate selected.  Each decision in this \emph{Risky Rate, Risky Day} treatment thus has a 10\% implementation probability.

The \emph{Certain Rate (CR)} treatment dimension eliminates risk regarding the rate to be implemented.  In this treatment, subjects are informed that $R_2=1.25$ will certainly be implemented; decisions for all prices $R_i\ne1.25$ are hypothetical, which I exclude from my analysis.

The \emph{Certain Day (CD)} treatment dimension eliminates risk regarding the day from which a decision is selected.  I inform subjects in this treatment that I will reveal the randomly-selected day \emph{before} their day-two decisions. Accordingly, the day to be implemented is risky for \emph{all} subjects on day zero, but certain for subjects with CD treatment on day two. Half of the subjects with this treatment learn that their day-two decisions are hypothetical; I exclude all hypothetical decisions from analysis and compensate by doubling the sample size of the CD dimension.\footnote{I considered this alternate design: If day zero is selected, inform after day-two decisions; if day two is selected, inform before day-two decisions.  I rejected this design because subjects would lack complete prior information about the timing of the resolution of risk.}

\Cref{tbl:treat-implement-prob} shows decision implementation probability by treatment cell $T$.

\begin{table}\centering
  \begin{threeparttable}
    \caption{Probability of decision implementation \label{tbl:treat-implement-prob}}
    \begin{tabular}{lcc}
    \toprule
    & \multicolumn{2}{c}{Decision chosen on} \\
    \cmidrule{2-3}
    Treatment & day $d=0$ & day $d=2$ \\
    \midrule
    Baseline & $1/10$ & $1/10$ \\
    Certain Day & $1/10$ & $1/5$ \\
    Certain Rate & $1/2$ & $1/2$ \\
    Certain Rate, Certain Day & $1/2$ & $1$ \\
    \bottomrule
    \end{tabular}
    \begin{tablenotes} \footnotesize
        \emph{Note:} Probabilities of implementation of the effort allocation choice $e_{2,d}$ (chosen on decision-day $d$ at rate $R_2=1.25$).
    \end{tablenotes}
  \end{threeparttable}
\end{table}

My primary interest is the interaction of Certain Rate and Certain Day treatments.  On day zero subjects know that their choice at $R_2=1.25$ made on either day zero or made on day two will be selected with certainty.  Then on day two, prior to making a decision, subjects learn from which day a decision will be selected.  Subjects who learn day two is selected thus make a decision on day two that is certainly implemented.  That is, these subjects choose their impending same-day effort level, knowing this choice will be implemented with certainty, with the tasks due imminently.  I hypothesize that present bias is more pronounced under certainty than under risk.

\subsection{Interface}

I carefully designed the interface to bolster subjects' understanding of the choice process and the implementation mechanism.  The interface guides every subject through a complete practice round at the beginning of each session before the real decisions and tasks.

\paragraph{Day selection} The interface shows each subject a list of their practice choices made on each decision-day (\cref{fig:ui-mechanism}). Upon clicking the button to proceed, the page visualizes random selection between the two decision-days by alternately highlighting the lists in quick succession before the highlight settles on one day as being selected.  Every subject simulates two practice coin-flips: the first trial selects the alternate decision-day, then the second selects the present day. The remainder of the practice round uses the choices made in the present session.

\paragraph{Rate selection} The interface next shows each subject the five practice choices made on the present day, arranged in a table similar to the juxtaposed allocation page (\cref{fig:ui-allocation-b}). Subjects with Certain Rate treatment see row four permanently highlighted and a reminder that only chocies at this rate will be implemented. Other subjects see no highlight at first, but upon clicking the button to proceed, a roulette-wheel sequence highlights each row quickly in succession. After traversing the table twice, the highlight settles on a randomly-selected decision.

With a practice allocation selected, subjects view a practice task interface that requests the corresponding amount of work to be completed on the present day.  Subjects then exit the practice round and begin an identical sequence with real decisions and tasks.

\section{Model and methodology}

Assuming quasi-hyperbolic discounting, within-day power-function effort costs, and background effort of $\omega$, on each decision-day $d=0$ and $d=2$ the decision-maker optimizes%
\begin{equation}\min_{e_{i,d}^t} \beta^{\mathbb{1}(d=0)} \left( e_{i,d}^2 + \omega \right) ^\alpha + \beta \delta^7 \left( e_{i,d}^9 + \omega \right) ^\alpha, \text{ subject to } e_i^2 + R_i e_i^9 = 360,\label{eq:dm-problem}\end{equation} choosing effort $e_{i,d}^t$ for each price ratio $R_i \in \left \{ 0.5, 0.75, 1, 1.25, 1.5 \right \} $ and workday $t \in \{2,9\}$. This model uses $\delta$ as a daily discount factor, while $\beta$ discounts future-day effort. Assuming the independence axiom, the resultant intertemporal Euler equation is:%
\begin{equation}\left( \frac{e_{i,d}^2+\omega}{e_{i,d}^9+\omega} \right)^{\alpha-1} = \frac{\beta^{\mathbb{1}(d=2)} \delta^7}{R_i}\label{eq:euler}\end{equation}
\parencite{Lawrance1991}. Logarithms linearize this equation as%
\begin{equation}\underbrace{\vphantom{\frac{e_{i,d}^2}{e_{i,d}^9}} \ln \frac{e_{i,d}^2+\omega}{e_{i,d}^9+\omega}}_{\eqqcolon E_{i,d}} = \underbrace{\vphantom{\frac{e_{i,d}^2}{e_{i,d}^9}} \frac{\ln \delta}{\alpha - 1} }_{\eqqcolon \theta_\textrm{delay}} 7 + \underbrace{\vphantom{\frac{e_{i,d}^2}{e_{i,d}^9}} \frac{-1}{\alpha - 1} }_{\eqqcolon \theta_\textrm{lnrate}} \ln R_i + \underbrace{\vphantom{\frac{e_{i,d}^2}{e_{i,d}^9}} \frac{\ln \beta}{\alpha - 1} }_{\eqqcolon \theta_\textrm{present}} \underbrace{\vphantom{\frac{e_{i,d}^2}{e_{i,d}^9}} \mathbb{1}(d=2) }_{\eqqcolon \mathbb{1}(\textrm{pr})}. \label{eq:reduced-form-original}\end{equation} Let us define the variables as shown under braces in \cref{eq:reduced-form-original} to simplify notation.  An additive error term produces an estimatable reduced-form, with $s$ indexing subjects:%
\begin{equation}E_{i,d,s} = \theta_\textrm{delay} 7 + \theta_\textrm{lnrate} \ln R_i + \theta_\textrm{present} \mathbb{1}(\textrm{pr})_{d} + \varepsilon_{i,d,s}. \label{eq:reduced-form}\end{equation}

Let $\mathbb{1}(\textrm{cr})$ indicate Certain Rate treatment and $\mathbb{1}(\textrm{cd})$ Certain Day treatment. We interact the full-factorial of these with the present-workday indicator, $\mathbb{1}(d=2)$, to obtain an estimatable pooled reduced-form regression model:
\begin{align}
 E_{i,d,s} &= \theta_{\textrm{delay}} 7 + \theta_{\textrm{lnrate}} \ln R_i + \theta_{\textrm{present}} \mathbb{1}(d=2)_{d} + \theta_{\textrm{cr}} \mathbb{1}(\textrm{cr})_{s} \mathbb{1}(d=2)_{d} \notag\\
 & \quad + \theta_{\textrm{cd}} \mathbb{1}(\textrm{cd})_{s} \mathbb{1}(d=2)_{d} + \theta_{\textrm{cr,cd}} \mathbb{1}(\textrm{cr})_{s} \mathbb{1}(\textrm{cd})_{s} \mathbb{1}(d=2)_{d} + \varepsilon_{i,d,s}. \label{eq:reduced-form-pooled}
\end{align}
This specification allows recovery of $\beta_T$ that varies by treatment cell $T$ (the supplement offers details).

\subsection{Hypotheses}

My primary hypothesis is that an interaction exists between the immediacy effect and the certainty effect. That is, present bias at certainty differs from present bias with risk.  The baseline treatment with risk regarding the rate and risk regarding the decision-day is standard in the literature, here with each decision having an implementation probability of $1/10$.  When treated with Certain Rate and Certain Day, the day-two choice for $R_2=1.25$ is implemented with certainty (probability of one).

\begin{hypothesis}
Present-bias is more intense under implementation certainty (with both Certain Rate and Certain Day treatment) than when the decision involves both types of risk (decision-day and rate both unrealized), in which each decision has an implementation probability of $1/10$: $\beta_{\textrm{cr,cd}}<\beta$.
\end{hypothesis}

For completeness I further hypothesize that present bias at certainty differs from present bias with any uncertainty---that is, with only one dimension of risk.

\begin{hypothesis}
Present-bias is more intense under implementation certainty than when the decision involves rate risk but has decision-day certainty (implementation probability of $1/5$): $\beta_{\textrm{cr,cd}}<\beta_{\textrm{cd}}$.
\end{hypothesis}

\begin{hypothesis}
Present-bias is more intense under implementation certainty than when the decision involves decision-day risk but has rate certainty (implementation probability of $1/2$): $\beta_{\textrm{cr,cd}}<\beta_{\textrm{cr}}$.
\end{hypothesis}

I do not hypothesize further how the type of risk may matter.  For example, controlling for the implementation probability, perhaps risk regarding the rate is most influential, perhaps driven by the income effect of the price ratios.  Research regarding types of risk and underlying mechanisms is left to future research.

\subsection{Statistical methodology}

Each subject allocates 360 tasks (in day-two valuation) between day two and day nine at various price ratios.  Each subject must also complete ten mandatory real-effort tasks on each day.  A subject might most prefer a negative effort allocation to a workday (that is, a net gain of leisure on that day), which my environment does not facilitate: the environment bounds elicited day-two effort such that $e_{i,d}^2\in[0,360]$ for all rates $R_i$ on each decision-day $d$.  A two-limit Tobit model accommodates this censoring.

To estimate the model with a power cost function $c(e)\coloneqq(e+\omega)^\alpha$, we must specify some background effort $\omega>0$.  The primary analysis will use $\omega=10$ as the background effort, corresponding to the mandatory daily tasks.  Subjects may perform other tasks throughout the day that we might wish to include in $\omega$; my supplement offers results that demonstrate robustness to various background effort levels.

\subsection{Identification of present bias}

As represented by the factor $\beta$, present bias is identified from a two-day window. On Monday, day zero, I assume that subjects view both Wednesday, day two, and the next Wednesday, day nine, as part of the future. Then on day two, I assume that subjects view that same day as the present and continue to view day nine as the future.

One could reasonably argue that Monday and Wednesday of the same week may both feel relatively present, while the following week may feel relatively distant. This would imply that present bias would be better identified from a week-long delay, as in \textcite{AugenblickNiederleSprenger2015}. However, this is an empirical question, and \textcite{Augenblick2018} studies exactly how present bias varies with short delays. Using similar real-effort tasks, he finds that present bias quickly diminishes within three days, with two days capturing most present bias. In the present study, the use of a two-day window will yield conservative estimates of $\beta$ (biased upward). A week-long delay might more accurately identify present bias, but likely at the cost of greater attrition.  Regardless, in the present study I simply intend to use average treatment effects to test my hypotheses.

\subsection{Identification of discounting}

The daily discount factor $\delta$ exponentially discounts the future. In this environment, we identify $\delta$ (jointly with $\alpha$) using variation in the rate $R_i$.

Suppose that marginal cost of effort is constant within a day, so that the effort-cost convexity parameter $\alpha=1$. Then the ratio $\delta^7/R_i$ determines how a decision-maker allocates her workload between day two and day nine. If $\delta^7/R_i=1$, she is indifferent to how the workload is split; otherwise she will allocate the entire workload to a single day. For example, if she discounts the future ($\delta<1$) but she can trade day-two and day-nine work one-for-one (when $R_i=1$), she will choose to do all of the work on day nine.

Instead, assume that the decision-maker has an increasing marginal cost of within-day effort, so that $\alpha>1$. Then if $R_i=1$ and $\delta=1$, she would divide the workload evenly between day two and day nine. This is because she values smoothing effort between workdays, since additional within-day effort becomes more costly. Then as either the rate $R_i$ or the discount factor $\delta$ changes, the decision-maker will choose a different workload split between day two and day nine, balancing the benefit of smoothing effort against the inferior rate.

Because $\delta^7/R_i$ and effort-cost convexity $\alpha$ jointly explain allocation between workdays, $\alpha$ and $\delta$ are jointly identified by rate $R_i$ variation.

\subsection{Between- and within-subject identification}

My novel experimental design and identification strategy permits estimation of the present-bias factor $\beta$ with some subjects making only two incentivized decisions (the interaction of Certain Rate and Certain Day treatments).  This is possible because effort-cost convexity $\alpha$ and the discount factor $\delta$ are estimated with the pooled regression, relying on variation in rate $R_i$ from Risky Rate treatment. Meanwhile, $\beta_T$ is estimated between-subjects for each treatment cell $T$.\footnote{While within-subject estimation of the present-bias factor is possible, each point-estimate $\beta_s$ would rely on only two observations, $e_{1,0}$ and $e_{1,2}$, per subject $s$ given Certain Rate treatment.}

Other researchers who wish to evaluate present bias under certainty need not implement my full-factorial design.  My design will identify $\alpha$, $\delta$, and $\beta_T$ with only two treatment cells: a cell with Risky Rate and a cell with Certain Rate and Certain Day.  The cell with Risky Rate can use either Risky Day or Certain Day; however the latter requires a larger sample to achieve the same statistical power if the study excludes hypothetical decisions and avoids deception.  For my present study I implement a full-factorial design to investigate each dimension.

\section{Results}

I recruited subjects from an online piece-rate labor marketplace with an equivalent median hourly wage under \$5 and jobs that commonly involve transcribing invoices or tagging photographs \parencite{Newman2019}.\footnote{On Amazon Mechanical Turk I sampled residents of the United States and Canada with at least 1,000 jobs completed with 98\% approval.}  I described my ``multi-day counting project'' as three sessions with a combined 30--50 minutes of tasks which paid \$1.50 for each session and a \$5 bonus for completing all three.  Given that the experiment involves tasks similar to those typical of the marketplace, it is a \emph{framed field experiment} \parencite{HarrisonList2004}.

My instructions provided sample tasks, explained the task allocation process, and stressed the three dates of participation: Monday 28 October 2019 (day zero), Wednesday 30 October (day two), and Wednesday 6 November (day nine).  Consenting subjects answered an eight-question comprehension survey which paid \$1.50 regardless of the responses (the supplement provides all experimental instruments).

Of the 389 comprehension survey submissions, 220 provided informed consent and only correct responses; I enrolled these subjects in my experiment.  Of these 220 subjects, 206 (93.6\%) enrolled in and completed day-zero decisions.  From this first session to the last, sample attrition was only 26 of 206 subjects (12.6\%).\footnote{On day two, 192 subjects (93.2\%) returned and completed that day's decisions.  One of each subject's ten decisions was implemented, upon which 188 subjects (97.9\%) completed the selected day-two effort.  Finally, on day nine, 180 subjects (95.7\%) returned and completed the session, thus earning the completion bonus.} The median subject completed a combined 340 tasks in 36 minutes (with quartiles $q_1=28$ and $q_3=47$ minutes).

\subsection{Descriptive results}\label{sec:results-descriptive}

While log-effort-ratio is the correctly specified choice variable given the model in \cref{eq:dm-problem}, let us consider a more intuitive outcome: day-two effort-share $\varphi \coloneqq e_{i,d}^2 / 360$. Because the Certain Price treatment only incentivizes $R_2=1.25$, I only analyze choice data at this rate. Note that because day-nine effort is more productive than day-two effort at this rate, effort is split evenly between workdays when $e_{2,d}^2 = e_{2,d}^9 = 160$ and thus $\varphi = 160/360 = 0.44$.

\begin{figure}\centering
\caption{Histograms of effort-share chosen for day two at $R_2=1.25$ for each treatment \label{fig:hist-treatment}}
\includegraphics[width=\textwidth, trim=0cm 0cm 0cm 0cm, clip]{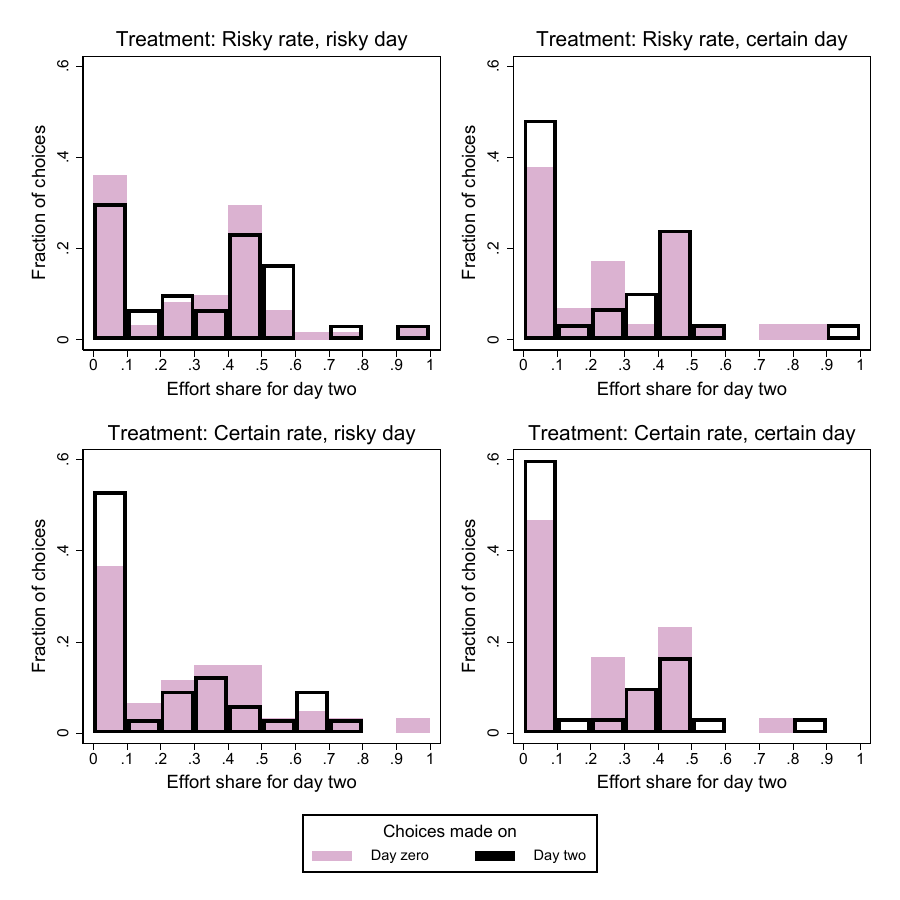}
\end{figure}

\Cref{fig:hist-treatment} offers histograms of effort-share choices by treatment.  Given an effort-share choice in advance of the workday (day zero), a smaller choice on the workday itself (day two) suggests present bias.  Thus this graph illustrates present bias if the outlined bars shift to the left of the solid bars.

We see a striking feature at 0.0--0.10: in treatments with at least one certainty treatment, many subjects choose a day-two allocation much different than their day-zero allocation.  On day two, with work imminent, they choose to exert 0.0--0.10 of effort on day two. Compare the filled bar against the outlined bar for the 0.0--0.10 bin in each treatment; more subjects choose an allocation in this bin on day two (when the work is imminent) relative to day zero (when the work is distant) with either certainty treatment.

This simple descriptive graph intuitively suggests evidence of present bias in at least some treatments.  We now turn to regression analysis to make use of all incentivized choice data.

\subsection{Regression results}

We now consider the regression results as presented in \cref{tbl:results-longitudinal-restr} and \cref{fig:beta-restr}.  We notably find no present bias in the baseline treatment. Meanwhile, relative to the baseline treatment with risk, present bias under certainty is vastly different economically.  With a point-estimate of 0.58 under certainty, subjects value the present 1.7 times as much as they value the future.

\begin{table}[p!]\centering
 \makebox[\textwidth][c]{
  \begin{threeparttable}
    \caption{Regression results\label{tbl:results-longitudinal-restr}}
    \begin{tabular}{ld{2.4}d{2.4}d{3.4}d{2.4}d{2.4}d{3.4}}
        \toprule
& & & \multicolumn{4}{c}{\(p\)-value of \(\chi^2_1\) test that parameter equals} \\
        \cmidrule{4-7}
Param. & \multicolumn{1}{c}{Estim.} & \multicolumn{1}{c}{Std.~Err.} & \multicolumn{1}{c}{$1$} & \multicolumn{1}{c}{$\beta$} & \multicolumn{1}{c}{$\beta_\mathrm{cd}$} & \multicolumn{1}{c}{$\beta_\mathrm{cr}$} \\
        \midrule
        $\beta_\mathrm{cr,cd}$& 0.581 & (0.108) & <0.001  & <0.001 & 0.005 & 0.458 \\
        $\beta_\mathrm{cr}$    & 0.679 & (0.109) &  0.003  &  0.011 & 0.046 &       \\
        $\beta_\mathrm{cd}$    & 0.921 & (0.057) &  0.166  &  0.327 &       &       \\
        $\beta$                & 1.009 & (0.055) &  0.873  &        &       &       \\
        \midrule
        $\delta$               & 0.986 & (0.004) &  0.001  &        &       &       \\
        $\alpha$               & 1.282 & (0.045) & <0.001  &        &       &       \\
        \bottomrule
     \end{tabular}
    \begin{tablenotes}
        \footnotesize \emph{Note:}
        $897$ observations ($161$ left- and $95$ right-censored) from $180$ subjects.
        Robust standard errors in parentheses are clustered on subject using a two-limit Tobit model.
    \end{tablenotes}
  \end{threeparttable}
 }
\end{table}

\begin{figure}[p!]\centering
  \caption{Regression estimates of $\beta_T$\label{fig:beta-restr}}
  \makebox[\textwidth][c]{
\begin{tikzpicture}[domain=0:1,scale=1]
\begin{axis}[scale=1.75, ymin=0.3, ymax=1.2, xmin=0, xmax=1,
  ytick={0.3,0.4,...,1.1}, ytick align=outside, ytick pos=left,
  xtick={0,0.2,...,1}, xtick align=outside, xtick pos=left,
  xlabel=Implementation probability,
  ylabel={$\hat\beta_T$},
  axis line style={draw opacity=0.5},
  x tick label style={
                /pgf/number format/fixed,
                /pgf/number format/fixed zerofill,
                /pgf/number format/precision=1},
  y tick label style={
                /pgf/number format/fixed,
                /pgf/number format/fixed zerofill,
                /pgf/number format/precision=1},
  grid=major, clip=false]
\addplot+[no marks,black,dashed,line width=1.2pt] {1};
\addplot+[
  only marks,
  mark options={black, scale=1},
  visualization depends on=\thisrow{alignment} \as \alignment,
  point meta=explicit symbolic,
  nodes near coords,
  every node near coord/.style={anchor=\alignment, inner sep=1, outer sep=4},
  error bars/.cd, 
    y fixed,
    y dir=both, 
    y explicit,
    error bar style={width=4pt, line width=4pt, white!60!red}
] table [x=x, y=y, y error=error, col sep=comma, row sep=crcr, meta index=3] {
      x,         y, error,      label, alignment\\
    0.1, 1.008819, 0.10788506, \contour{white}{$1.009$},   0 \\
    0.2, 0.921123, 0.11172627, \contour{white}{$0.921$},   0 \\
    0.5, 0.679382, 0.21316274, \contour{white}{$0.679$},   0 \\
    1.0, 0.580839, 0.21232092, \contour{white}{$0.581$},   0 \\
};
\tikzset{labelnode/.style={font=\fontsize{10}{10}\selectfont},
bracelabelnode/.style={midway,yshift=-2.2ex}}
\node[labelnode,right,align=left] at ($(axis cs:0.110,1.110)$) {\contour{white}{Risky rate,}\\\contour{white}{risky day}};
\node[labelnode,right,align=left] at ($(axis cs:0.210,0.85)$) {\contour{white}{Risky rate,}\\\contour{white}{certain day}};
\node[labelnode,right,align=left] at ($(axis cs:0.510,0.85)$) {\contour{white}{Certain rate,}\\\contour{white}{risky day}};
\node[labelnode,left,align=right] at ($(axis cs:0.990,0.75)$) {\contour{white}{Certain rate,}\\\contour{white}{certain day}};
\node[labelnode,right,align=left, outer sep=2] at ($(axis cs:0.1, 1.008819)$) {\contour{white}{$\beta$}};
\node[labelnode,right,align=left, outer sep=2] at ($(axis cs:0.2, 0.9211234)$) {\contour{white}{$\beta_\textrm{cd}$}};
\node[labelnode,right,align=left, outer sep=2] at ($(axis cs:0.5, 0.6793819)$) {\contour{white}{$\beta_\textrm{cr}$}};
\node[labelnode,right,align=left, outer sep=2] at ($(axis cs:1.0, 0.5808388)$) {\contour{white}{$\beta_\textrm{cr,cd}$}};
\draw [decorate,decoration={brace,mirror},scale=0.2,black,thick]
  (0.5,1.9) -- (1.0,1.9) node[bracelabelnode]{\contour{white}{$p=0.327$}};
\draw [decorate,decoration={brace,mirror},scale=0.2,black,thick]
  (1.0,1.6) -- (2.5,1.6) node[bracelabelnode]{\contour{white}{$p=0.046$}};
\draw [decorate,decoration={brace,mirror},scale=0.2,black,thick]
  (0.5,1.3) -- (2.5,1.3) node[bracelabelnode]{\contour{white}{$p=0.011$}};
\draw [decorate,decoration={brace,mirror},scale=0.2,black,thick]
  (2.5,1.2) -- (5.0,1.2) node[bracelabelnode]{\contour{white}{$p=0.458$}};
\draw [decorate,decoration={brace,mirror},scale=0.2,black,thick]
  (1.0,0.9) -- (5.0,0.9) node[bracelabelnode]{\contour{white}{$p=0.005$}};
\draw [decorate,decoration={brace,mirror},scale=0.2,black,thick]
  (0.5,0.6) -- (5.0,0.6) node[bracelabelnode]{\contour{white}{$p<0.001$}};
\end{axis}
\end{tikzpicture}
  }
\end{figure}

We reject the null hypothesis that $\beta_\textrm{cr,cd}=\beta$ at $p<0.001$. This provides clear evidence that the introduction of a substantial amount of risk significantly moderates present bias.

We also find that addition of risk in the rate dimension alone also drastically moderates present bias relative to the baseline, rejecting $\beta_\textrm{cr,cd}=\beta_\textrm{cd}$ with $p=0.005$.  However we do not find a similar result in comparing certainty to the addition of risk in the decision-day dimension, failing to reject that $\beta_\textrm{cr,cd}=\beta_\textrm{cr}$ with $p=0.458$.  Recall that with Certain Rate treatment, we rely on only two observations per subject; otherwise we have ten observations per subject.  Indeed the standard errors for $\beta$ with Certain Rate treatment are roughly double those in the other treatments, suggesting that a larger sample may reveal significance.

Beyond concerns of statistical power, we can conjecture that the implementation probability plays an important role (\cref{fig:beta-restr}).  When making a day-two decision with Certain Rate and Risky Day treatments, subjects know that their present decision will be selected by one side of a coin flip.  This might make the decision sufficiently salient to preserve a high degree of present bias, especially with only a single incentivized decision on each day.  Regardless, the testing of such hypotheses and underlying mechanisms is left to future work.

\subsection{Additional considerations}

\minisec{Attrition bias}
While sample attrition (12.6\%) was remarkably low for an online experiment across ten days, we should look for evidence of selective attrition.  For example, Certain Rate treatment might have lower attrition as it guarantees an income effect, whereas the Risky Rate treatment does not in expectation.

Only four subjects completed day-two decisions but did not complete the implemented day-two effort level.  Two of these made only hypothetical decisions on day two and are thus excluded from the analysis.  Both remaining subjects had Certain Rate treatment.  We conclude that attrition during day two was orthogonal to rate resolution.

Attrition between day two and day nine of eight subjects was highly balanced across treatments and rate selection.

\minisec{Effort-cost curvature}
We reject the hypothesis that $\alpha\geq1$ with $p<0.001$, satisfying the second-order condition for \cref{eq:dm-problem}. The results are robust to background effort $\omega$ of greater orders of magnitude, appropriate for having already worked prior to the sessions (see the supplement).

\section{Conclusion}

This study of dynamic inconsistency in real-effort provision finds that risk diminishes the intensity of present bias.  This includes uncertainty that arises from random-implementation mechanisms popular among experimental economists.  My experiment varies the implementation mechanism, thereby altering the probability of decision implementation.  The novel design permits pooled estimation of the present-bias factor in each of four treatment cells, including one that implements a single decision with certainty.

The effect of certainty on present bias is striking.  Under certainty I estimate $\hat{\beta}_\mathrm{cr,cd}=0.581$, while the baseline treatment finds no significant present bias with a point-estimate of $\hat{\beta}=1.009$.  These results present a remarkable treatment effect: risk significantly moderates present bias.

While most other studies find present bias in real-effort, some do not \parencite{ImaiRutterCamerer2021}.
The replication study of \textcite{AugenblickNiederleSprenger2015} is similar to my baseline treatment, using the same implementation mechanism, one-week delay, and similar interest rates. They estimate $\hat{\beta}=0.892$ with $p=0.05$ for University of California students transcribing blurry Greek letters.  My subjects, being workers in an online marketplace, may have substituted effort in my session with effort in other jobs.  Nevertheless the treatment effect suggests that the implementation mechanism affects present bias.

Experiments that seek an accurate point-estimate of the present-bias factor should include a decision with substantial immediate and certain consequences.  If complementarities between consequences do not pose a serious concern, the experiment might reasonably implement multiple such decisions.

My findings underscore the importance of unifying theories of time and risk, notably \textcite{ChakrabortyHalevySaito2020} (see \cref{sec:background-risk-delay,note:risk-delay-papers}).  Conversely, in testing decision-theoretic models, researchers should mind their incentive mechanisms and use decisions implemented with certainty when appropriate.

Empirical work on tempting goods may require decisions with salient and certain consequences, a potentially-critical design element for any study employing experimental methods.  Such work might study models of self-control, the effectiveness of commitment devices, or any application that depends on present-biased preferences.

Uncertainty may interact with non-stationary time preferences, leading to different behavior in strategic interactions.  For example, in labor contracts firms may exploit present bias with (un)certainty regarding compensation, effective productivity, or job duration.

Clearly the field of behavioral economics has much yet to learn about present bias, temptation, and related interventions. Continual improvement of experimental methodology will aid this pursuit.

\vspace{1\baselineskip}

\spacing{1}
{\RaggedRight \printbibliography }

\end{document}


\begin{center}
{\large
\phantom{.}\\
\vspace{6pt}
\textbf{\LARGE
Temptation: Immediacy and certainty\\
\vspace{12pt}
Supplement
}\\
\vspace{24pt}
J.~Lucas Reddinger\\
\vspace{9pt}
26 August 2024\\
\vspace{18pt}
}
\end{center}

\tableofcontents

\section{Additional results}

\begin{itemize}[noitemsep]
\item \Cref{tbl:subjects}: Subjects by treatment
\item \Cref{tbl:background-tasks}: Regression results with various background effort
\item \Cref{tbl:between-subjects-covar}: Estimated covariance matrix for the primary regression
\end{itemize}

\section{Experimental instruments}

\begin{itemize}[noitemsep]
\item \Cref{tbl:timeline}: Experimental timeline
\item \Cref{tbl:test}: Comprehension test
\end{itemize}

The interface included animations to convey the random selection procedure to subjects.  Only one slide from each animation is included in this document.

\begin{figure}[p]
\fbox{\includegraphics[width=\textwidth,trim=0 170.25cm 0 0cm,clip,page=1]{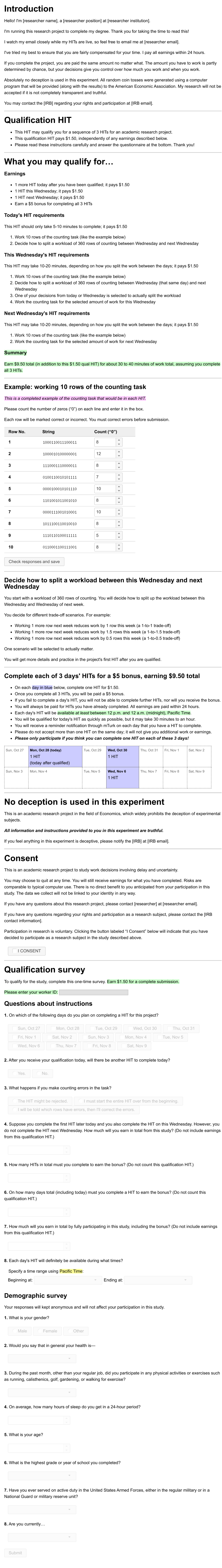}}
\caption{Qualification session instructions (1 of 6)}
\label{fig:ui:qual-1}
\end{figure}

\begin{figure}[p]
\fbox{\includegraphics[width=\textwidth,trim=0 69cm 0 0cm,clip]{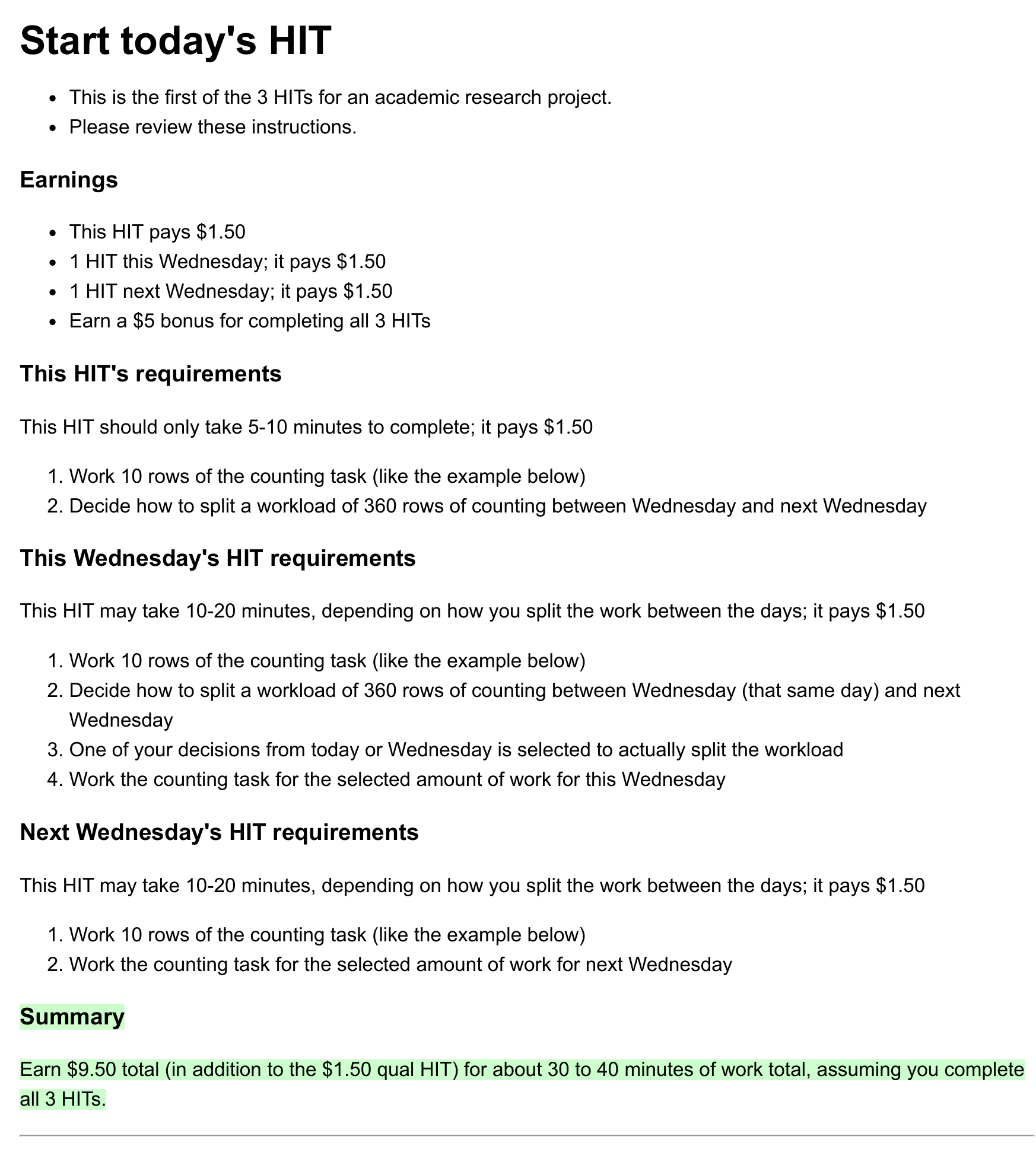}}
\caption{Day zero session instructions (1 of 3)}
\label{fig:ui:day0-start-1}
\end{figure}

\begin{figure}[p]
\fbox{\includegraphics[width=\textwidth,trim=0 7cm 0 0,clip,page=5]{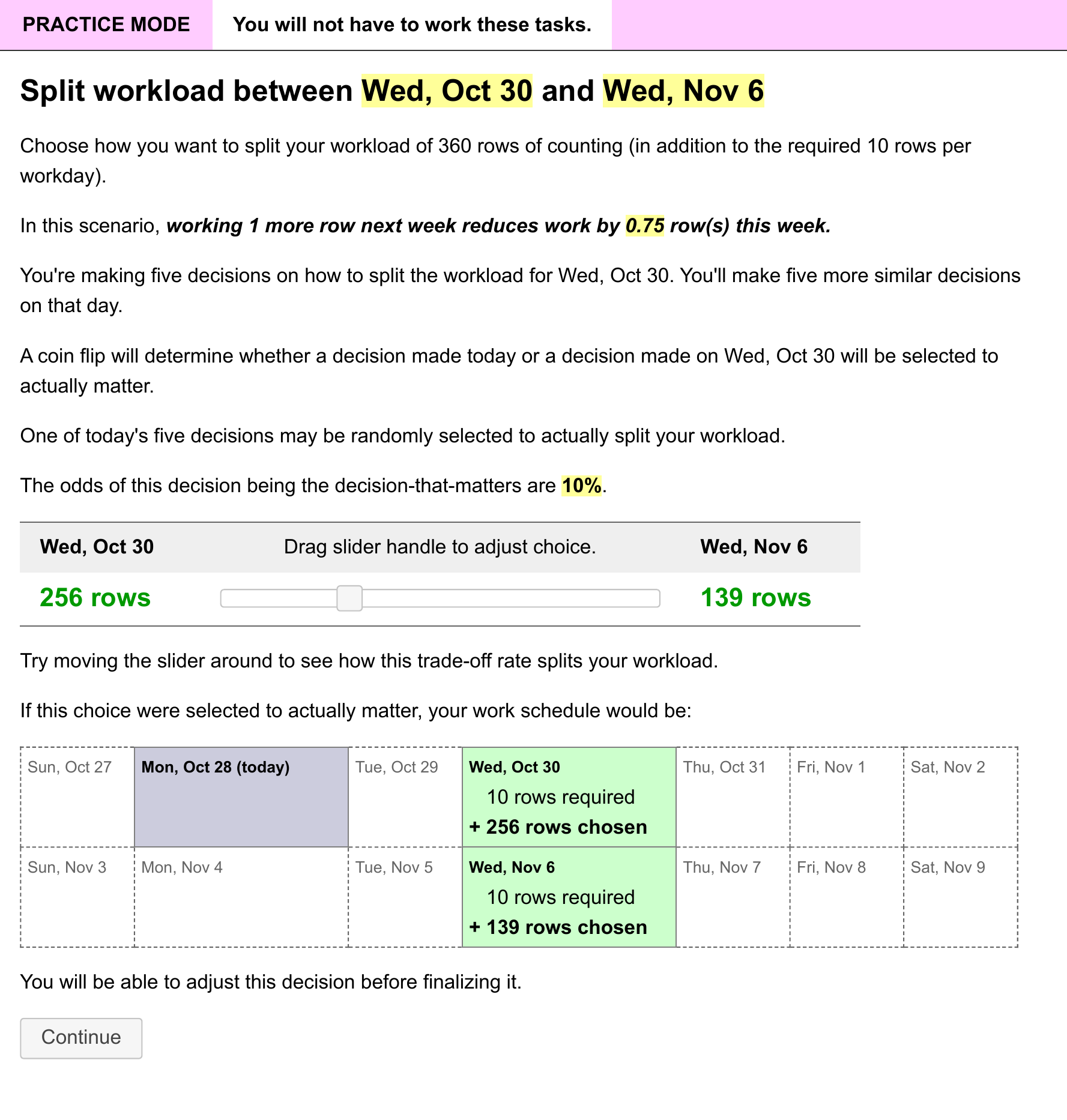}}
\caption[Task allocation, separate]{Task allocation, separate: This interface allows the subject to allocate their workload between days.}
\label{fig:ui:task-allocation-a}
\end{figure}

\begin{figure}[p]
\fbox{\includegraphics[width=\textwidth,trim=0 14cm 0 0,clip,page=17]{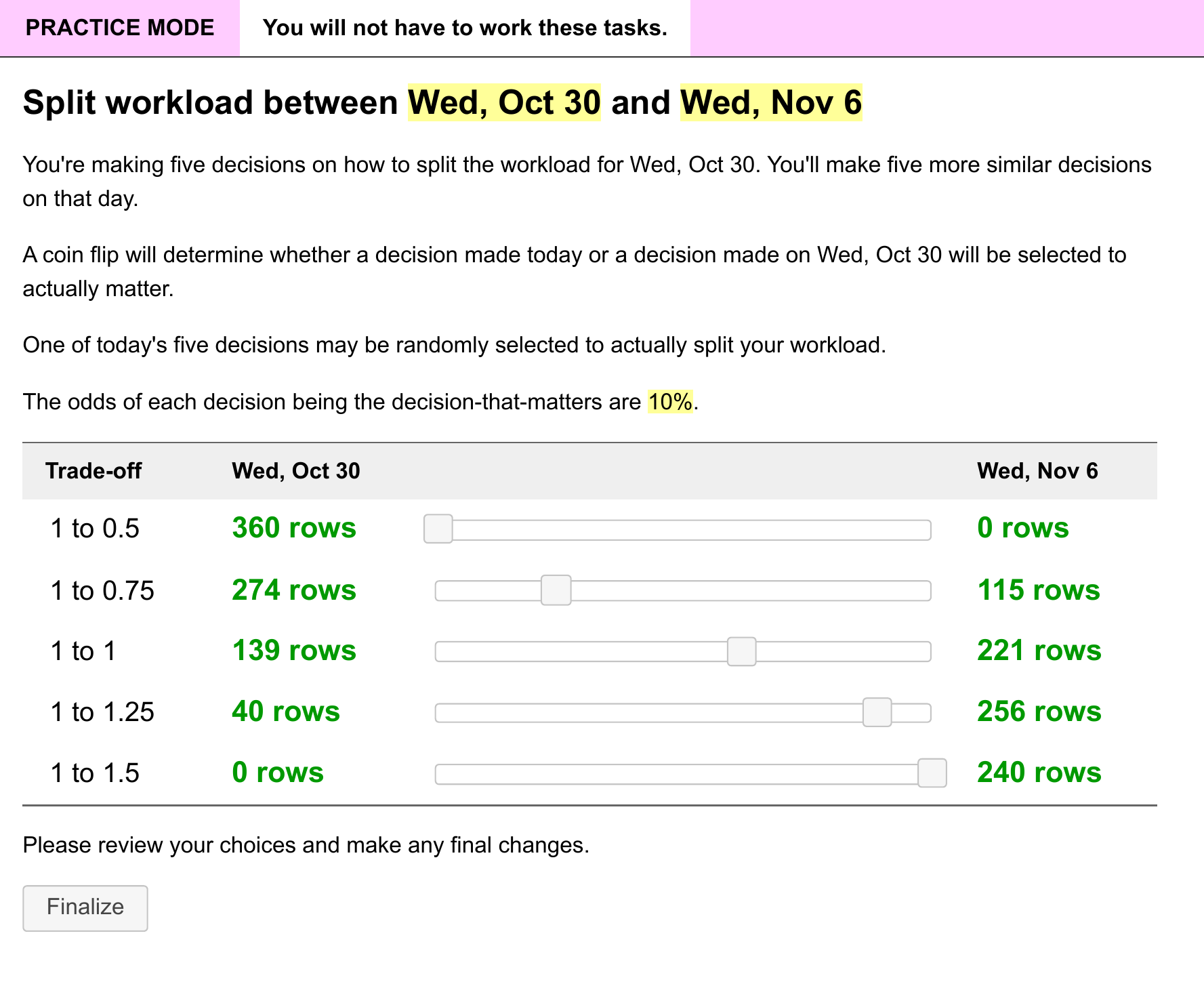}}
\caption[Task allocation, juxtaposed]{Task allocation, juxtaposed: This interface allows the subject to allocate their workload between days.}
\label{fig:ui:task-allocation-b}
\end{figure}

\begin{figure}[p]
\fbox{\includegraphics[width=\textwidth,trim=0 19cm 0 0,clip,page=1]{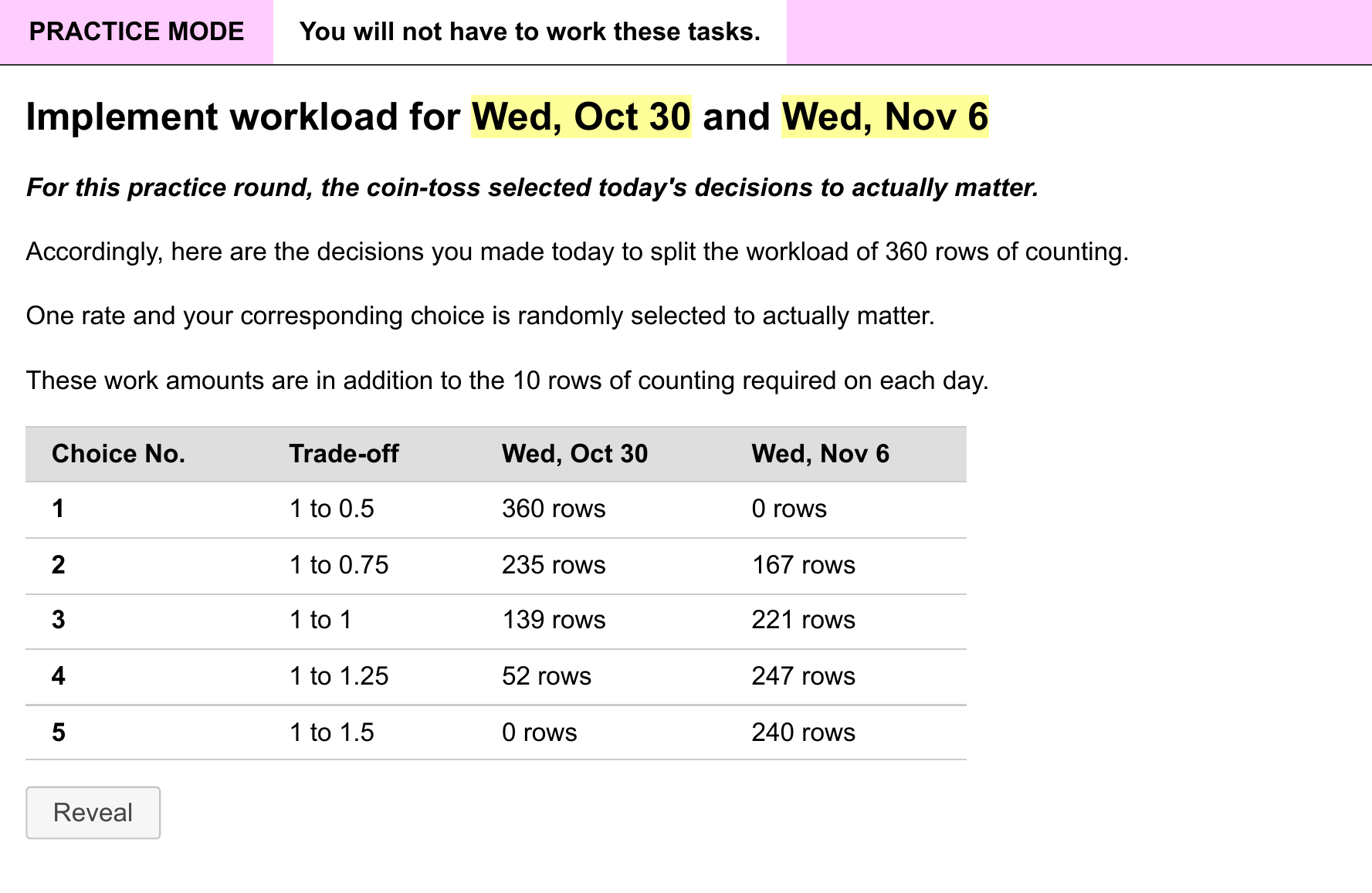}}
\caption[Rate selection]{This interface gives the subject intuition regarding the selection procedure between rates.}
\label{fig:ui:rate-selection}
\end{figure}

\begin{figure}[p]
\fbox{\includegraphics[width=\textwidth,trim=0 11cm 0 0,clip,page=1]{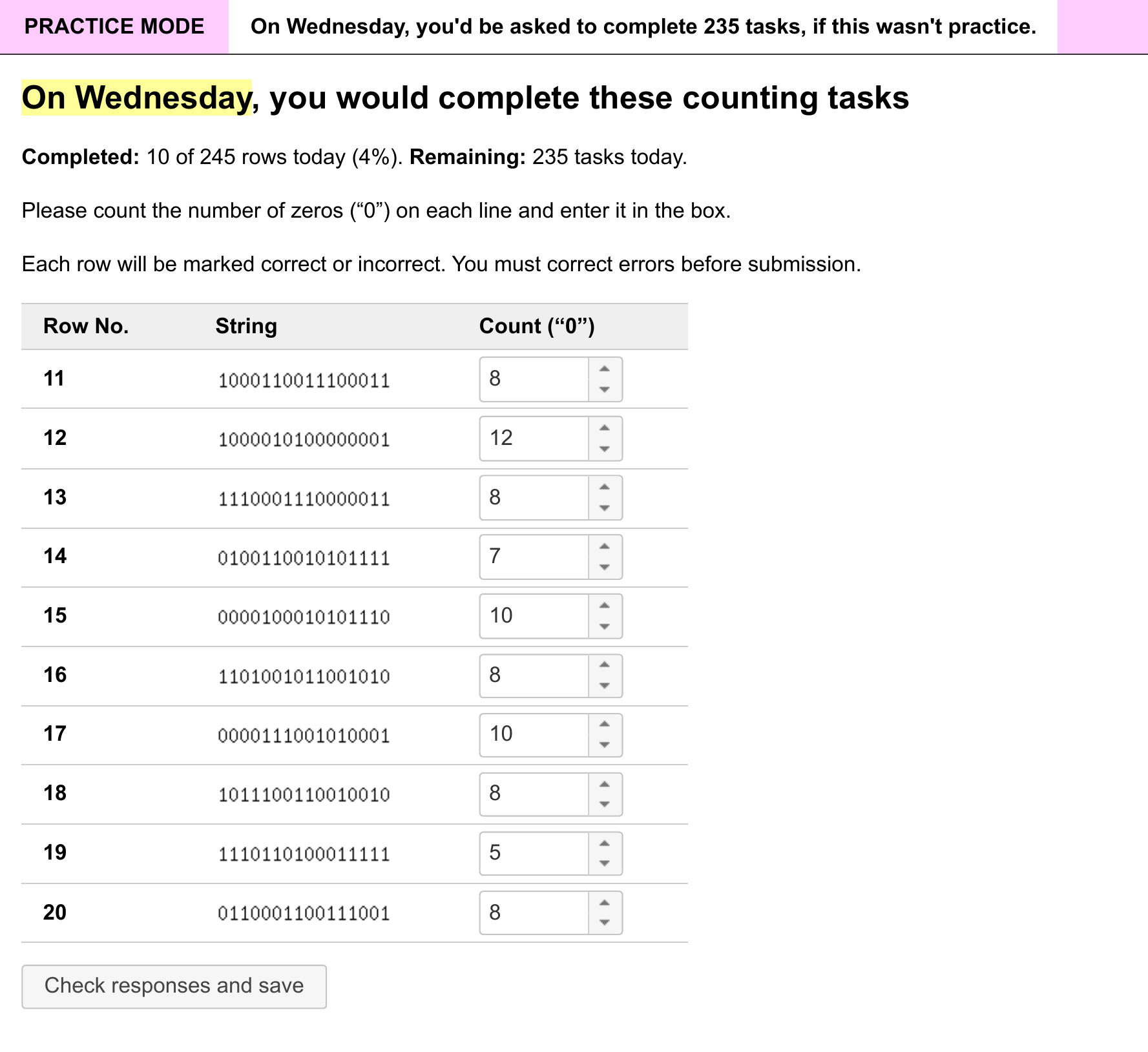}}
\caption[Example of implemented tasks]{This interface shows the subject examples of the implemented tasks.}
\label{fig:ui:implemented-tasks}
\end{figure}

\begin{itemize}[noitemsep]
\item \Cref{fig:ui:qual-1,fig:ui:qual-2,fig:ui:qual-3,fig:ui:qual-4,fig:ui:qual-5,fig:ui:qual-6}: Qualification session instructions
\item \Cref{fig:ui:day0-start-1,fig:ui:day0-start-2,fig:ui:day0-start-3}: Day zero session instructions
\item \Cref{fig:ui:tasks-reqd}: Required tasks
\item \Cref{fig:ui:task-allocation-a}: Task allocation, separate
\item \Cref{fig:ui:task-allocation-b}: Task allocation, juxtaposed
\item \Cref{fig:ui:day-selection}: Decision-day selection
\item \Cref{fig:ui:rate-selection}: Rate selection
\item \Cref{fig:ui:implemented-tasks}: Example of implemented tasks
\end{itemize}

\section{Parameter recovery}

Parameters in the model of the main text are recovered as follow:
\begin{align*}
  \beta_{\textrm{}} &= \exp \frac{\theta_{\textrm{present}}}{-\theta_{\textrm{lnrate}}}, & \beta_{\textrm{cr}} &= \exp \frac{\theta_{\textrm{present}} + \theta_{\textrm{cr}}}{-\theta_{\textrm{lnrate}}}, \\
  \beta_{\textrm{cd}} &= \exp \frac{\theta_{\textrm{present}} + \theta_{\textrm{cd}}}{-\theta_{\textrm{lnrate}}}, & \beta_{\textrm{cr,cd}} &= \exp \frac{\theta_{\textrm{present}} + \theta_{\textrm{cr}} + \theta_{\textrm{cd}} + \theta_{\textrm{cr,cd}}}{-\theta_{\textrm{lnrate}}}, \\
  \delta_{\textrm{}} &= \exp \frac{\theta_{\textrm{delay}}}{-\theta_{\textrm{lnrate}}}, \textrm{ and} & \alpha &= 1-\theta_{\textrm{lnrate}}^{-1}.
\end{align*}

\begin{table}[t]\centering
\def\sym#1{\ifmmode^{#1}\else\(^{#1}\)\fi}
\begin{threeparttable}
\caption{Subject counts by treatment\label{tbl:subjects}}
\begin{tabular}{*5{l}*3{d{3.0}}}
    \toprule
    \multicolumn{5}{c}{Treatment} & \multicolumn{3}{c}{Subject count on day} \\
    \cmidrule(r){1-5} \cmidrule(r){6-8}
No. & Day & Rate & Selects day\tnote{1} & Rate order\tnote{2} & \multicolumn{1}{c}{Zero} & \multicolumn{1}{c}{Two} & \multicolumn{1}{c}{Nine} \\
    \midrule
  1 & Risky   & Risky   & Two  & Original &  8 &  8 &  8 \\
  2 & Risky   & Certain & Two  & Original & 10 & 10 & 10 \\
  3 & Risky   & Risky   & Zero & Original &  9 &  8 &  7 \\
  4 & Risky   & Certain & Zero & Original &  8 &  7 &  7 \\
  5 & Certain & Risky   & Two  & Original & 17 & 16 & 15 \\
  6 & Certain & Certain & Two  & Original & 17 & 16 & 15 \\
  7 & Certain & Risky   & Zero & Original & 18 & 16 & 16 \\
  8 & Certain & Certain & Zero & Original & 15 & 15 & 13 \\
    \midrule
  9 & Risky   & Risky   & Two  & Reversed &  9 &  8 &  7 \\
 10 & Risky   & Certain & Two  & Reversed &  8 &  8 &  7 \\
 11 & Risky   & Risky   & Zero & Reversed &  9 &  8 &  8 \\
 12 & Risky   & Certain & Zero & Reversed &  9 &  9 &  8 \\
 13 & Certain & Risky   & Two  & Reversed & 17 & 15 & 14 \\
 14 & Certain & Certain & Two  & Reversed & 17 & 16 & 15 \\
 15 & Certain & Risky   & Zero & Reversed & 18 & 16 & 15 \\
 16 & Certain & Certain & Zero & Reversed & 17 & 16 & 15 \\
    \midrule
    \multicolumn{5}{l}{Total subject count by day} & 206 & 192 & 180 \\
    \bottomrule
    \end{tabular}
    \begin{tablenotes}[para,flushleft] \footnotesize
      \item[1] The day from which a decision was selected for implementation.
      \item[2] Decisions are presented in the order of $\mathcal{R}$ or its reverse.
    \end{tablenotes}
\end{threeparttable}

\end{table}

\begin{table}\centering
\def\sym#1{\ifmmode^{#1}\else\(^{#1}\)\fi}
\begin{threeparttable}
\caption{Regression results with various background effort\label{tbl:background-tasks}}
\begin{tabularx}{5in}{X*{5}{d{3.5}}}
\toprule
                    &\multicolumn{4}{c}{Background effort, \(\omega\)}  \\\cmidrule{2-5}
                    &\multicolumn{1}{c}{ \( 10 \) }&\multicolumn{1}{c}{ \( 1200 \) }&\multicolumn{1}{c}{ \( 2400 \) }&\multicolumn{1}{c}{ \( 4800 \) }\\
\midrule
Equivalent duration of \(\omega\)\tnote{\dag}&\multicolumn{1}{c}{1 minute}&\multicolumn{1}{c}{2 hours}&\multicolumn{1}{c}{4 hours}&\multicolumn{1}{c}{8 hours}\\
\midrule
\(\beta\)           &       1.009&       0.984&       0.982&       0.981\\
                    &     (0.055)&     (0.068)&     (0.069)&     (0.069)\\
\(\beta_\mathrm{cd}\)&       0.921&       0.884&       0.882&       0.880\\
                    &     (0.057)&     (0.072)&     (0.074)&     (0.075)\\
\(\beta_\mathrm{cr}\)&       0.679&       0.665&       0.662&       0.661\\
                    &     (0.109)&     (0.113)&     (0.114)&     (0.114)\\
\(\beta_\mathrm{cr,cd}\)&       0.581&       0.569&       0.566&       0.565\\
                    &     (0.108)&     (0.113)&     (0.114)&     (0.114)\\
\(\delta\)          &       0.986&       0.987&       0.987&       0.987\\
                    &     (0.004)&     (0.005)&     (0.005)&     (0.005)\\
\(\alpha\)          &       1.282&       5.048&       8.668&      15.898\\
                    &     (0.045)&     (0.688)&     (1.317)&     (2.575)\\
\midrule
\multicolumn{2}{l}{\(p\)-value of \(\chi^2_1\) test:}\\ \quad\(\beta=\beta_\mathrm{cd}\)&       0.327&       0.366&       0.369&       0.371\\
\quad\(\beta=\beta_\mathrm{cr}\)&       0.011&       0.019&       0.019&       0.020\\
\quad\(\beta=\beta_\mathrm{cr,cd}\)&       0.001&       0.002&       0.002&       0.002\\
\quad\(\beta_\mathrm{cd}=\beta_\mathrm{cr}\)&       0.046&       0.084&       0.086&       0.087\\
\quad\(\beta_\mathrm{cr,cd}=\beta_\mathrm{cd}\)&       0.005&       0.012&       0.012&       0.012\\
\quad\(\beta_\mathrm{cr,cd}=\beta_\mathrm{cr}\)&       0.458&       0.481&       0.481&       0.481\\
\quad\(\beta=1\)    &       0.873&       0.808&       0.795&       0.788\\
\quad\(\beta_\mathrm{cd}=1\)&       0.166&       0.109&       0.108&       0.108\\
\quad\(\beta_\mathrm{cr}=1\)&       0.003&       0.003&       0.003&       0.003\\
\quad\(\beta_\mathrm{cr,cd}=1\)&      <0.001&      <0.001&      <0.001&      <0.001\\
\quad\(\delta=1\)   &       0.001&       0.014&       0.017&       0.019\\
\quad\(\alpha=1\)   &      <0.001&      <0.001&      <0.001&      <0.001\\
\bottomrule
\end{tabularx}
\begin{tablenotes}[para,flushleft] \footnotesize \emph{Note:} 897 observations (161 left- and 95 right-censored) from 180 subjects. Robust standard errors in parentheses are clustered on subject using a two-limit Tobit model. Excludes subjects who did not complete all sessions. The \emph{ex ante} specification uses background effort \(\omega\!=\!10\). \item[\dag] The median duration of 10 real-effort tasks is 1 minute.
\end{tablenotes}
\end{threeparttable}

\end{table}

\begin{table}\centering
\def\sym#1{\ifmmode^{#1}\else\(^{#1}\)\fi}
\begin{threeparttable}
\caption{Estimated covariance matrix for the primary regression with \(\omega=10\)\label{tbl:between-subjects-covar}}
\begin{tabular}{l*{7}{d{2.5}}}
\toprule
            &         \multicolumn{1}{c}{\(\beta\)}&         \multicolumn{1}{c}{\(\beta_\mathrm{cd}\)}&         \multicolumn{1}{c}{\(\beta_\mathrm{cr}\)}&         \multicolumn{1}{c}{\(\beta_\mathrm{cr,cd}\)}&         \multicolumn{1}{c}{\(\delta\)}&         \multicolumn{1}{c}{\(\alpha\)}\\
\midrule
\(\beta\)   &     0.00303&    -0.00086&    -0.00099&    -0.00090&     0.00006&     0.00014\\
\(\beta_\mathrm{cd}\)&    -0.00086&     0.00325&     0.00020&     0.00027&    -0.00002&    -0.00044\\
\(\beta_\mathrm{cr}\)&    -0.00099&     0.00020&     0.01183&     0.00297&     0.00011&    -0.00244\\
\(\beta_\mathrm{cr,cd}\)&    -0.00090&     0.00027&     0.00297&     0.01173&     0.00012&    -0.00263\\
\(\delta\)  &     0.00006&    -0.00002&     0.00011&     0.00012&     0.00002&    -0.00008\\
\(\alpha\)  &     0.00014&    -0.00044&    -0.00244&    -0.00263&    -0.00008&     0.00206\\
\bottomrule
\end{tabular}
\end{threeparttable}

\end{table}

\begin{table}\centering
\def\sym#1{\ifmmode^{#1}\else\(^{#1}\)\fi}
\begin{threeparttable}
\caption{Experimental timeline\label{tbl:timeline}}
    \begin{tabular}{lrlr}
    \toprule
    Day zero & \multicolumn{2}{l}{``Qualification HIT''} & Payment of \$1.50 within twenty-four hours \\[6pt]
          & 1. & \multicolumn{2}{l}{Instructions} \\
          & 2. & \multicolumn{2}{l}{Consent} \\
          & 3. & \multicolumn{2}{l}{Comprehension test} \\[6pt]
    \multicolumn{4}{c}{\emph{A subject qualifies for the next session if and only if all comprehension answers are correct.}} \\
    \midrule
    Day zero & \multicolumn{2}{l}{``Monday's HIT''} & Payment of \$1.50 within twenty-four hours \\[6pt]
          &  1. & \multicolumn{2}{l}{Instructions} \\
          &  2. & \multicolumn{2}{l}{\emph{Practice:} Ten mandatory tasks that would need to be completed} \\
          &  3. & \multicolumn{2}{l}{\emph{Practice:} Effort allocation between day two and day nine, presented separately} \\
          &  4. & \multicolumn{2}{l}{\emph{Practice:} Effort allocation between day two and day nine, presented juxtaposed} \\
          &  5. & \multicolumn{2}{l}{\emph{Practice:} How today's decisions are used (resolution of decision-day risk)} \\
          &  6. & \multicolumn{2}{l}{\emph{Practice:} How today's decisions are used (resolution of rate risk)} \\
          &  7. & \multicolumn{2}{l}{\emph{Practice:} View implemented tasks that would need to be completed} \\
          &  8. & \multicolumn{2}{l}{Complete the ten mandatory tasks} \\
          &  9. & \multicolumn{2}{l}{Effort allocation between day two and day nine, presented separately} \\
          & 10. & \multicolumn{2}{l}{Effort allocation between day two and day nine, presented juxtaposed} \\[6pt]
    \multicolumn{4}{c}{\emph{A subject qualifies for the next session if and only if all parts of this session are completed.}} \\
    \midrule
    Day two & \multicolumn{2}{l}{``This Wednesday's HIT''} & Payment of \$1.50 within twenty-four hours \\[6pt]
          &  1. & \multicolumn{2}{l}{Instructions} \\
          &  2. & \multicolumn{2}{l}{\emph{Practice:} Ten mandatory tasks that would need to be completed} \\
          &  3. & \multicolumn{2}{l}{\emph{Practice:} Effort allocation between day two and day nine, presented separately} \\
          &  4. & \multicolumn{2}{l}{\emph{Practice:} Effort allocation between day two and day nine, presented juxtaposed} \\
          &  5. & \multicolumn{2}{l}{\emph{Practice:} How today's decisions are used (resolution of decision-day risk)} \\
          &  6. & \multicolumn{2}{l}{\emph{Practice:} How today's decisions are used (resolution of rate risk)} \\
          &  7. & \multicolumn{2}{l}{\emph{Practice:} View implemented tasks that would need to be completed} \\
          &  8. & \multicolumn{2}{l}{Complete the ten mandatory tasks} \\
          &  9. & \multicolumn{2}{l}{\emph{Certain Day treatment only:} One day is selected for implementation} \\
          & 10. & \multicolumn{2}{l}{Effort allocation between day two and day nine, presented separately} \\
          & 11. & \multicolumn{2}{l}{Effort allocation between day two and day nine, presented juxtaposed} \\
          & 12. & \multicolumn{2}{l}{\emph{Risky Day treatment only:} One day is selected for implementation} \\
          & 13. & \multicolumn{2}{l}{\emph{Risky Rate treatment only:} One rate is selected for implementation} \\
          & 14. & \multicolumn{2}{l}{Complete the implemented tasks for today} \\[6pt]
    \multicolumn{4}{c}{\emph{A subject qualifies for the next session if and only if all parts of this session are completed.}} \\
    \midrule
    Day nine & \multicolumn{2}{l}{``Next Wednesday's HIT''} & Payment of \$6.50 within twenty-four hours \\[6pt]
           & 1. & \multicolumn{2}{l}{Instructions} \\
           & 2. & \multicolumn{2}{l}{Complete the ten mandatory tasks} \\
           & 3. & \multicolumn{2}{l}{Complete the implemented tasks for today} \\
    \bottomrule
    \end{tabular}
    \begin{tablenotes}[para,flushleft] \footnotesize
      \emph{Note:} In the labor marketplace used, \emph{HIT} is common nomenclature for a single job.
    \end{tablenotes}
\end{threeparttable}

\end{table}

\begin{table}\centering
\def\sym#1{\ifmmode^{#1}\else\(^{#1}\)\fi}
\begin{threeparttable}
\caption{Comprehension test\label{tbl:test}}
    \begin{tabular}{p{0.25in}p{5.5in}}
    \toprule
    1. & On which of the following days do you plan on completing a HIT for this project? \\
       & \emph{One checkbox per date:} \\
       & \hspace{0.5cm} Sun, Oct 27; Mon, Oct 28; Tue, Oct 29; Wed, Oct 30; Thu, Oct 31; Fri, Nov 1; \\
       & \hspace{0.5cm} Sat, Nov 2; Sun, Nov 3; Mon, Nov 4; Tue, Nov 5; Wed, Nov 6; Thu, Nov 7; \\
       & \hspace{0.5cm} Fri, Nov 8; Sat, Nov 9 \\
    \midrule
    2. & After you receive your qualification today, will there be another HIT to complete today? \\
       & \emph{Radio buttons:} Yes; No \\
    \midrule
    3. & What happens if you make counting errors in the task? \\
       & \emph{Radio buttons:} \\
       & \hspace{0.5cm} The HIT might be rejected. \\
       & \hspace{0.5cm} I must start the entire HIT over from the beginning. \\
       & \hspace{0.5cm} I will be told which rows have errors, then I'll correct the errors. \\
    \midrule
    4. & Suppose you complete the first HIT later today and you also complete the HIT on this Wednesday. However, you do not complete the HIT next Wednesday. How much will you earn in total from this study? (Do not include earnings from this qualification HIT.) \\
       & \emph{Text input formatted as USD currency} \\
    \midrule
    5. & How many HITs in total must you complete to earn the bonus? (Do not count this qualification HIT.) \\
       & \emph{Text input} \\
    \midrule
    6. & On how many days total (including today) must you complete a HIT to earn the bonus? (Do not count this qualification HIT.) \\
       & \emph{Text input} \\
    \midrule
    7. & How much will you earn in total by fully participating in this study, including the bonus? (Do not include earnings from this qualification HIT.) \\
       & \emph{Text input formatted as USD currency} \\
    \midrule
    8. & Each day's HIT will definitely be available during what times? Specify a time range using \emph{Pacific Time}: \\
       & \emph{Selection menu:} \\
       & \hspace{0.5cm} Beginning at: 07:00 am; 08:00 am; 09:00 am; 10:00 am; 11:00 am; 12:00 pm; \\
       & \hspace{1cm}   01:00 pm; 02:00 pm; 03:00 pm; 04:00 pm; 05:00 pm; 06:00 pm \\
       & \hspace{0.5cm} Ending at: 12:00 pm; 1:00 pm; 2:00 pm; 3:00 pm; 4:00 pm; 5:00 pm; 6:00 pm; \\
       & \hspace{1cm}   7:00 pm; 8:00 pm; 9:00 pm; 10:00 pm; 11:00 pm; 12:00 am (midnight); \\
       & \hspace{1cm}   01:00 am; 02:00 am \\
    \bottomrule
    \end{tabular}
    \begin{tablenotes}[para,flushleft] \footnotesize
      \emph{Note:} In the labor marketplace used, \emph{HIT} is common nomenclature for a single job.
    \end{tablenotes}
\end{threeparttable}

\end{table}

\begin{figure}[p]
\fbox{\includegraphics[width=\textwidth,trim=0 127.5cm 0 46.75cm,clip,page=1]{figure_interface_0_qualify}}
\caption{Qualification session instructions (2 of 6)}
\label{fig:ui:qual-2}
\end{figure}

\begin{figure}[p]
\fbox{\includegraphics[width=\textwidth,trim=0 89cm 0 81.5cm,clip,page=1]{figure_interface_0_qualify}}
\caption{Qualification session instructions (3 of 6)}
\label{fig:ui:qual-3}
\end{figure}

\begin{figure}[p]
\fbox{\includegraphics[width=\textwidth,trim=0 51.25cm 0 120cm,clip,page=1]{figure_interface_0_qualify}}
\caption{Qualification session instructions (4 of 6)}
\label{fig:ui:qual-4}
\end{figure}

\begin{figure}[p]
\fbox{\includegraphics[width=\textwidth,trim=0 15cm 0 158cm,clip,page=1]{figure_interface_0_qualify}}
\caption{Qualification session instructions (5 of 6)}
\label{fig:ui:qual-5}
\end{figure}

\begin{figure}[p]
\fbox{\includegraphics[width=\textwidth,trim=0 0.5cm 0 194cm,clip,page=1]{figure_interface_0_qualify}}
\caption{Qualification session instructions (6 of 6)}
\label{fig:ui:qual-6}
\end{figure}

\begin{figure}[p]
\fbox{\includegraphics[width=\textwidth,trim=0 31cm 0 33cm,clip]{figure_interface_1_start}}
\caption{Day zero session instructions (2 of 3)}
\label{fig:ui:day0-start-2}
\end{figure}

\begin{figure}[p]
\fbox{\includegraphics[width=\textwidth,trim=0 0.5cm 0 71.5cm,clip]{figure_interface_1_start}}
\caption{Day zero session instructions (3 of 3)}
\label{fig:ui:day0-start-3}
\end{figure}

\begin{figure}[p]
\fbox{\includegraphics[width=\textwidth,trim=0 12cm 0 0,clip,page=1]{figure_interface_2_demo_task_reqd}}
\caption[Required tasks]{Required tasks: This interface shows the subject examples of the task.}
\label{fig:ui:tasks-reqd}
\end{figure}

\begin{figure}[p]
\fbox{\includegraphics[width=\textwidth,page=1]{figure_interface_4_demo_decision}}
\caption[Decision-day selection]{This interface gives the subject intuition regarding the selection procedure between days.}
\label{fig:ui:day-selection}
\end{figure}

